\documentstyle[aipmod,eqalign,amssymb,graphicx,times]{article}
\advance \textwidth by 1in
\advance \oddsidemargin by -0.5in
\advance \textheight by 1in
\advance \topmargin by -0.5in
 \font\tensfb=cmssbx10 \relax
\def\dvv{d^3{\bf v}}
\def\dvu{d^3{\bf u}}
\def\abs#1{\left|#1\right|}\def\ave#1{\left<#1\right>}
\def\sigin{\Sigma_{\rm in}}\def\sigout{\Sigma_{\rm out}}
\def\sign{\mathop{\rm sign}\nolimits}
\def\mat#1{\hbox{\tensfb #1}}
\def\eqref#1{Eq.~(\ref{#1})}\def\Eqref#1{Equation~(\ref{#1})}
\def\figref#1{Fig.~\ref{#1}}
\def\tabref#1{Table~\ref{#1}}
\def\Eqlab#1{\eqn(\eqlab{#1})}
\def\refref#1{Ref.~\citenum{#1}}
\let\grapprox=\gtrsim
\let\lsapprox=\lesssim

\begin{document}
\date{PPPL--2222 (June 1985)\\
Phys.\ Fluids {\bf 29}(1), 180--192 (Jan.~1986)}

\title{Current in Wave Driven Plasmas}
\author{Charles F. F. Karney and Nathaniel J. Fisch\\
Plasma Physics Laboratory, Princeton University\\Princeton, NJ 08544}
\maketitle
\begin{abstract}

A theory for the generation of current in a toroidal plasma by
radio-frequency waves is presented.  The effect of an opposing electric
field is included, allowing the case of time varying currents to be
studied.  The key quantities that characterize this regime are
identified and numerically calculated.
Circuit equations suitable for use in ray-tracing and
transport codes are given.

\end{abstract}
\section{INTRODUCTION}\label{intro}

In recent years there has been considerable interest in generating
steady-state currents in a plasma with rf waves.  In particular, it was
predicted \cite{Fisch1} that these currents could be efficiently
generated by waves whose phase velocities are several times the
electron thermal speed.  This prediction has been confirmed by numerous
experiments in which the current was driven by lower-hybrid waves.
These results allow us to contemplate a steady-state tokamak reactor in
which the toroidal current is driven by lower-hybrid waves.  This is an
attractive proposition not only because of the advantages inherent in
steady-state operation (less thermal stress, higher duty cycle, etc.),
but also because it opens up the possibility that the ohmic winding of
the tokamak can be eliminated entirely, leading to a cheaper and more
compact reactor.  This latter possibility can be realized if rf waves
are successful not only in sustaining the plasma current ``steady-state
current drive,'' but also in increasing the plasma current ``rf current
ramp-up.''  In fact, experiments have demonstrated that this too is
possible.  From a theoretical point of view, the important additional
ingredient in these experiments is the dc electric field which opposes
the increase of the plasma current.  The electric field is also present
in schemes where the rf is used to recharge the transformer at constant
current.

Recently, we presented a theory for rf current drive in the presence of
an electric field \cite{ramp1}.  This theory predicted that rf energy
could be efficiently converted to poloidal field energy if the wave
phase velocity were approximately equal to the electron runaway
velocity.  This theory has been compared \cite{ramp2} with data from the
{\sc plt} experiment \cite{Jobes}, and excellent agreement is found.  In
\refref{ramp1}, the linearized Boltzmann equation was approximately
solved by integrating the corresponding Langevin equations using a
Monte-Carlo method.  In this paper, we use a more elegant theory to
calculate the efficiency of the current ramp-up based on an adjoint
formulation for the Boltzmann equation \cite{Fischa}.  Although the
limits of validity of this theory are the same as for \refref{ramp1},
this theory is more amenable to accurate evaluation on a computer, and
it is more easily extended to include effects which are omitted here.

Let us begin by reiterating the physical picture given in
\refref{ramp1}.  Consider an electron travelling in the positive
direction at several times the thermal speed and which has just absorbed an
incremental amount of rf energy.  Suppose there is an electric field
tending to decelerate this electron.  The question is:  Where does this
incremental energy end up?  If the electron is slow compared to the
runaway velocity, the electron slows down primarily due to collisions
and so the rf energy goes to bulk heating.  On the other hand, if the
electron is fast, the electron is slowed down by the electric field.
In this case, the rf energy is coupled to the plasma circuit and
appears as poloidal field energy.  Unfortunately, fast electrons have a
high probability of pitch-angle scattering into the reverse direction
and running away.  A runaway electron drains energy out of the electric
field, leading to a degradation of the ramp-up efficiency.  However,
there is a window around the runaway velocity where the electrons are
slowed down principally by the electric field and yet where the probability
of running away is very small.  This is the favorable regime in which rf
energy can be efficiently converted to poloidal field energy.

From the foregoing discussion, we see that two ingredients are needed
for an accurate theoretical treatment of this problem.  First, the
electric field must be treated as large.  In the efficient regime, the
force on the electron due to the electric field must be comparable to
that due to collisions.  Second, a two-dimensional treatment is
required.  Analyses based on a one-dimensional Fokker--Planck equation
do not predict the important physical phenomenon of rf-generated
reverse runaways.

We briefly review the history of theoretical studies of current drive
in the presence of an electric field.  The earliest theoretical studies
\cite{Fisch1} of lower-hybrid current drive assumed that there was no
electric field in the plasma.  This is the appropriate limit for a
steady-state reactor.  However, some of the early experiments conducted
to verify the predictions of the theory were conducted in regimes where
the ohmic electric field was still present.  This prompted a series of
papers \cite{Harveya,Liu,Muschietti,An,Appert} dealing with rf current
drive in the presence of an assisting electric field (i.e., the
electric field and the rf both accelerate the electrons in the same
direction).  The principal focus of these papers was the calculation of
an enhanced runaway rate when the phase velocity of the waves is in the
neighborhood of the runaway velocity.  An opposing electric field was
treated by Borrass and Nocentini \cite{Borrass} within the framework of
a one-dimensional Fokker--Planck analysis.  A similar model has been
employed more recently by Liu {\it et al} \cite{Liu-ramp}.
As noted above, such an analysis cannot include rf-generated
reverse runaways.  Also due to the crudity of
the one-dimensional equation, the results are only accurate to within a
factor of order unity even when no runaways are generated.  A
two-dimensional treatment of the problem in the small electric field
limit has been given by Start \cite{Start} for the case of current
drive by electron-cyclotron waves.  This work neglected
electron-electron collisions.  This defect was removed and the results
generalized to arbitrary current-drive methods by Fisch
\cite{Fisch-cond} employing an adjoint formulation.  This work yields
accurate results when the electric field is small.  However, the results
are inapplicable in the regime of efficient ramp-up where the phase
velocity is comparable to the runaway velocity.  Our earlier paper
\cite{ramp1} was the first to combine a two-dimensional treatment with
a large electric field.  This paper allowed an accurate calculation of
the ramp-up efficiency in cases of practical interest, and identified
the regime in which high efficiencies can be expected.  The present work
is a continuation and expansion of that earlier paper.  Besides using
the more sophisticated method \cite{Fischa} for solving the Boltzmann
equation, we endeavor to give the results in a form that allows both easy
comparison with experiments and easy implementation within the
framework of ray-tracing or transport codes.

The paper is organized as follows:  We begin with the linearized
Boltzmann equation for the perturbed electron distribution in the
presence of an electric field and an rf source (Sec.~\ref{basic}).  Some
approximations and normalizations are made to reduce this equation to a
more manageable form.  The use of the adjoint method \cite{Fischa} for
solving the resulting equation is described (Sec.~\ref{formulation}).
Next (Sec.~\ref{results}) the adjoint equation is solved numerically to
give the runaway probability and the Green's function for the current.
The latter quantity is reduced to a simple form which involves just two
functions of velocity.  An expression for the total current density is
given (Sec.~\ref{circuit}) and this is put into a form which is
easy to calculate.  How the rf-driven current interacts with the
electric field to produce poloidal field energy is considered
(Sec.~\ref{applications}), and the results are applied to experiments.

\section{BOLTZMANN EQUATION}\label{basic}

Consider a uniform electron-ion plasma, initially at equilibrium.  For
$t>0$, it is subject to an electric field ${\bf E}(t)$ and a
wave-induced flux ${\bf S}({\bf v},t)$.  We will take the ions to be
infinitely massive, so that they form a stationary background off which
the electrons collide.  If the electric field and the wave-induced flux
are weak enough, the electron distribution remains close to a
Maxwellian for ${\cal E} \lsapprox T$ where $\cal E$ is the energy of an
electron ${1\over 2}mv^2$.  Substituting $f=f_m+f_1$ into the Boltzmann
equation for the electron distribution $f$ and linearizing then gives
	$$
{\partial\over\partial t}f_1+
{q{\bf E}(t)\over m}\cdot{\partial\over\partial \bf v}f_1
-C(f_1)=-{\partial\over\partial \bf v}\cdot{\bf S}
-{q{\bf E}(t)\over m}\cdot{\partial\over\partial \bf v}f_m
-\biggl[{\dot n\over n}+\Bigl({{\cal E}\over T}-{3\over 2}\Bigr)
{\dot T\over T}\biggr]f_m,
\Eqlab{lin} $$
	where
	$$f_m=n\Bigl({m\over2\pi T}\Bigr)^{3/2}\exp(-{\cal E}/T),$$
	and $$C(f)=C(f,f_m)+C(f_m,f)+C(f,f_i)$$
	is the linearized collision operator.  Here $q$, $m$, $n$, and
$T$ are the electron charge, mass, number density, and temperature.
Note that $q$ carries the sign of the electron charge (i.e., $q=-e$).

This equation is to be solved with initial condition
$f_1({\bf v},t=0)=0$.
We demand that the subsequent evolution of $f_1$ be such that
it be orthogonal to $1$ and ${\cal E}$, i.e., that it
have zero density and energy.  The zero-density condition is
satisfied with $\dot n=0$ since all the terms in \eqref{lin} are
particle conserving.  The zero-energy condition gives an equation for
the time evolution of $T$
$$
{3\over 2}n{dT\over dt}=\int m{\bf S}\cdot{\bf v}\,\dvv+
{\bf E}\cdot\int q{\bf v}f_1\,\dvv.
$$
The two terms on the right-hand side represent the heating due to the
waves and due to ${\bf E}\cdot{\bf J}$.

We now make three simplifying assumptions:  we assume that $f_1$ is
azimuthally symmetric about the ambient magnetic field; we take the
electric field to be constant and in the direction parallel to the
magnetic field ${\bf E}=E{\bf \hat v}_\parallel$; and we restrict our
attention to those cases where $\bf S$ is only finite where $v\gg v_t$,
where $v_t^2=T/m$ is the thermal velocity.  We may then solve
\eqref{lin} using the high-velocity form for $C$:
	$$C(f)=\Gamma\biggl[{1\over v^2}{\partial\over \partial v}f
+{1+Z\over 2v^3}
{\partial\over\partial \mu}(1-\mu^2){\partial\over\partial \mu}f\biggr],$$
	where $\mu=v_\parallel/v$,
$\Gamma=nq^4\ln\Lambda/4\pi\epsilon_0^2m^2$, $\epsilon_0$ is the
dielectric constant of free space, $\ln\Lambda$ is the Coulomb
logarithm, and $Z$ is the effective ion charge state.  We have included
pitch-angle scattering and frictional slowing down, but ignored energy
diffusion.  In the problem of steady-state current drive
\cite{relativity}, the energy diffusion term introduces corrections of
order $(v_t/v)^2$.  Another term neglected in this approximate collision
operator is the effect of the Maxwellian colliding off the perturbed
distribution, $C(f_m,f)$.  The
corrections due to this term \cite{relativity} are of
order $(v_t/v)^3$.  With these approximations, the collision operator does
not depend on the electron temperature $T$.  Formally, we may derive the
form for $C$ by taking $T\rightarrow0$.  In this limit, we have
$f_m\rightarrow n\delta({\bf v})$.

It is convenient to introduce some normalizations.  The runaway velocity
$v_r$ is that velocity at which collisional frictional force equals the
acceleration due to the electric field
	$$v_r\equiv-\sign(qE)\sqrt{m\Gamma/\abs{qE}}.$$
	Notice that the sign of $v_r$ is opposite to the direction in which
electrons run away.  The Dreicer velocity \cite{Dreicer} is given by
$-\sqrt{2+Z}\,v_r$.  Similarly, we define a runaway collision frequency
	$$\nu_r\equiv\Gamma/\abs{v_r}^3.$$
	The normalized time and velocity are given by
	$$\tau=\nu_r t$$ and $${\bf u}={\bf v}/v_r.$$
	The components of $\bf v$ have to be normalized with care:
$u_\perp=v_\perp/\abs{v_r}$, $u_\parallel=v_\parallel/v_r$, and
$u=v/\abs{v_r}$.  This implies that
$u_\parallel/u=\sign(v_r)v_\parallel/v$ so that the conversion of the
pitch-angle variable $\mu$ in $\bf v$ space to that in $\bf u$ space
involves multiplication by $\sign(v_r)$.
Other quantities are normalized in a similar way; however, we
shall use the same symbols as for the unnormalized quantities.  Thus, the
distribution functions $f_1$ and $f_m$ are normalized to $n/\abs{v_r}^3$, the
rf-induced flux $\bf S$ to $n\nu_rv_r/\abs{v_r}^3$, etc.
Under this normalization,
\eqref{lin} then becomes
	$${\partial\over\partial \tau}f_1+D(f_1)
=
-{\partial\over\partial \bf u}\cdot{\bf S},\eqn(\eqlab{linb})
$$
	with $f_1({\bf u},\tau=0)=0$ and with the operator $D$ defined
by
	$$D\equiv
-{\partial\over\partial u_\parallel}
-{1\over u^2}{\partial\over \partial u}
-{1+Z\over 2u^3}
{\partial\over\partial \mu}(1-\mu^2){\partial\over\partial \mu}.
$$
	Equation (\ref{linb}) is singular at the origin.  However,
because $f_1$ in \eqref{linb} describes a physical particle
distribution, it obeys a particle conservation law near the origin.  We
therefore require that, close to $u=0$, $f_1({\bf u}) = N(\tau)
\delta({\bf u})$ with $N(0)=0$ and
	$${dN\over d\tau}=\lim_{u\rightarrow0}
	\int_{-1}^1 2\pi f_1(u,\mu)\,d\mu.$$
	Equation (\ref{linb}) depends only on a single parameter $Z$.
The dependence on the electric field $E$ can be normalized away, since
the electric field defines the only natural velocity scale in the
problem, $v_r$.

\Eqref{linb} is amenable to various methods of solution, and it is
instructive to review these before describing the method used here.  The
most straightforward approach is to integrate \eqref{linb} directly on a
computer.  This method allows $\bf S$ to be determined directly in terms
of the electron distribution $f$.  However, a thorough understanding of
the problem requires that many different forms of $\bf S$ be used.
Therefore, this procedure is costly because the several parameters used
to specify $\bf S$ must be scanned.  This is essentially the method
adopted in the early numerical studies of steady-state current drive by
lower-hybrid waves \cite{Karney-lh}.

The situation is improved to some extent by noting that \eqref{linb} is
a linear equation for $f_1$.  It may be solved in terms of a Green's
function $g$ given by the equation
	$$\biggl({\partial\over\partial\tau}+D\biggr)g({\bf u},\tau;{\bf u}')
=0\Eqlab{green}$$
	with $g({\bf u},\tau=0;{\bf u}')=\delta({\bf u}-{\bf u}')$.  The
electron distribution is then given by the convolution
	$$f_1({\bf u},\tau)=
\int_0^\tau d\tau\int\dvu'\,{\bf S}({\bf u}',\tau')\cdot
{\partial\over\partial {\bf u}'}g({\bf u},\tau-\tau';{\bf u}').$$
	This approach reduces the problem to the determination of a
single function $g$ of two vector arguments ($\bf u$ and ${\bf u}'$)
and a one scalar argument ($\tau$).  However this is still a daunting
computational task.

A closely related technique is to formulate the problem as a set of
Langevin equations \cite{Wang},
	$$\eqalign{{du\over d\tau}&=-{1\over u^2}-\mu,\cr
{d\mu\over d\tau}&=A(\tau)-{1-\mu^2\over u},\cr}\Eqlab{monte}$$
	where the pitch-angle scattering is represented by the
stochastic term $A(\tau)$.  Assuming that $u(\tau)=u$ and
$\mu(\tau)=\mu$ are given (i.e., non-stochastic), then $A(\tau)$
satisfies
	$$\eqalign{\ave{A(\tau)}&=-{1+Z\over u^3}\mu,\cr
\ave{A(\tau)A(\tau')}&={1+Z\over u^3}(1-\mu^2)\delta(\tau-\tau'),\cr}
\Eqlab{aprop}$$
	where the angle brackets denote averaging over the ensemble
defined by all the realizations of $A$.  Consider following a particular
electron using Eqs.~(\ref{monte}).  Suppose that the electron is
observed to travel with velocity ${\bf u}'$ at $\tau=0$.  Then $g({\bf
u},\tau;{\bf u}')\,\dvu$ is the probability that the velocity of the
electron at time $\tau$ is in the volume element $\dvu$ located at $\bf
u$.  In Appendix~\ref{fokker}, it is shown that this conditional probability
$g$ satisfies \eqref{green}.  Thus the solution to \eqref{green} can be
found by determining the distribution of a large number of electrons
obeying Eqs.~(\ref{monte}) with initial conditions with ${\bf
u}(\tau=0)={\bf u}'$.  Consequently, moments of $g$ can be determined by
ensemble averages of the Langevin variables.  For example, the current
given by \eqref{green} may be found by
	$$\int\dvu\, u\mu\,g(u,\mu,\tau;u',\mu')
	=\ave{u(\tau)\mu(\tau)}.$$

Equations~(\ref{monte}) may be integrated numerically by noting that
	$$\int_\tau^{\tau+\Delta\tau}
A(\tau')\,d\tau'$$
	should be picked from an ensemble with mean
$-(1+Z)\mu\Delta\tau/u^3$ and variance
$(1+Z)(1-\mu^2)\Delta\tau/u^3$ where $u$ and
$\mu$ are the values of those variables at time $\tau$.  As long as
$\Delta\tau$ is sufficiently small, further details about the
distribution of $A$ are unimportant.

Now Eqs.~(\ref{monte}) are the equations solved in our earlier
paper \cite{ramp1}.  This shows the exact equivalence between the
approach adopted there and that employed in the present
work.  Because the Langevin equations describe the electron behavior in
a slightly more physical manner, they often help in the interpretation
of the solutions to the Boltzmann equation.  This is especially true
when some electrons run away.  The Langevin equations are also very easy
to solve numerically by a Monte-Carlo method (as was done in
\refref{ramp1}), although their solution tends to be much more costly
than just solving \eqref{green} directly.

Equations (\ref{monte}) however may be easily solved analytically in
the limit $u\rightarrow0$ (this is equivalent to taking the limit
$E\rightarrow 0$).  Taking an ensemble average of the equations, we obtain
	$$\eqalignno{{du\over d\tau}&=-{1\over u^2},\cr
{d\ave\mu\over d\tau}&=-{1+Z\over u^3}\ave\mu.
\cr}$$
	Note that $u$ is not a stochastic variable in this limit.
Consequently, the hierarchy of moment equations may be closed at this
point.  These are the slowing-down equations solved by Fisch and Boozer
\cite{Fisch-Boozer} to give the current moment of electron distribution
$\ave{u\mu}=u\ave\mu$.  This shows that the approach
used in that paper is equivalent to solving the Boltzmann equation.

\section{ADJOINT METHOD}\label{formulation}

The methods for solving the Boltzmann equation described in the
previous section all entail a large amount of computation.  (An
exception is the limit $E\rightarrow0$, when the ensemble-averaged
Langevin equations can
be solved analytically \cite{Fisch-Boozer}.)  The problem
with these methods is that they are all capable of giving the electron
distribution function $f_1$.  Since, in many cases, we are only
interested in specific moments of $f_1$, we may hope to reduce the
computational requirements substantially by using a method that
gives only those specific moments.  Suppose we wish to determine a particular
moment of $f_1$, namely
	$$H(\tau)=\int\dvu\,h_0({\bf u})f_1({\bf u},\tau).$$
	(For instance, the current density would be given by $h_0({\bf
u})=u_\parallel$.)  Let us define the corresponding moment of the
Green's function
	$$h({\bf u}',\tau)=\int\dvu\,h_0({\bf u})g({\bf u},\tau;{\bf u}').
\Eqlab{greena}$$
The moment $H$ is then given in terms of $h$ by
	$$H(\tau)=
\int_0^\tau d\tau'\int\dvu\,{\bf S}({\bf u},\tau')\cdot
{\partial\over\partial\bf u}h({\bf u},\tau-\tau').\Eqlab{mainb}$$

What is needed is some method of calculating $h$ which doesn't involve
having to find $g$.  This is provided by the adjoint formulation of
Fisch \cite{Fischa}.  He shows that $h({\bf u},\tau)$ satisfies
	$${\partial\over\partial \tau}h+D^*(h)
=0,\eqn(\eqlab{adjointb})
$$
	with $h({\bf u},\tau=0)=h_0({\bf u})$ and with the operator
$D^*$ defined by
	$$D^*\equiv
{\partial\over\partial u_\parallel}
+{1\over u^2}{\partial\over \partial u}
-{1+Z\over 2u^3}
{\partial\over\partial \mu}(1-\mu^2){\partial\over\partial \mu}.
$$
	The singularity at the origin is handled by the boundary
condition $h({\bf u}=0,\tau)=0$.  The operators $D$ and $D^*$ are
adjoint operators, so that
	$$\int \bigl(hD(f)-fD^*(h)\bigr) \,\dvu=0$$
	for all $f({\bf u})$ and $h({\bf u})$ satisifying $f({\bf
u}\rightarrow\infty)=0$ and $h({\bf u}=0)=0$.

Similar techniques were introduced earlier by Antonsen and Chu
\cite{Antonsen} and by Taguchi \cite{Taguchi} for the study of
steady-state current drive.  The significant improvements afforded by
\refref{Fischa} are the ability to determine arbitrary moments of $f_1$
and the inclusion of the time-dependence of $f_1$.  Both of these are
important in the problem of current ramp-up.

From the relation between the two Green's functions $h$ and $g$, we see
that $h$ has a simple interpretation.  Equations~(\ref{green}) and
(\ref{monte}) describe the evolution of a group of electrons released
at $\tau=0$ at velocity ${\bf u}'$.  Let us suppose that we are
interested in the current density so that $h_0({\bf u})=u_\parallel$.
Then $h({\bf u}',\tau)$ gives the mean current carried by those
electrons a time $\tau$ later.  How \eqref{adjointb} works is easily
seen by taking $u\gg1$ so that the electron only experiences the
electric field.  In the Boltzmann equation, the electrons have slowed
down to ${\bf u}'-\tau{\bf \hat u}_\parallel$ at time $\tau$.
Correspondingly in the adjoint equation, the initial condition
$h_0$ is transported in the reverse direction so that $h({\bf
u}',\tau) = h_0({\bf u}'-\tau{\bf \hat u}_\parallel)$.  Thus at time
$\tau$, we are provided with information about the electrons in their
current location.

Solving the Boltzmann equation by means of the adjoint formulation
results in a great simplification of the problem.  The adjoint equation
(\ref{adjointb}) is an equation of equal complexity to the original
Boltzmann equation (\ref{linb}).  However,  by solving
\eqref{adjointb} for a particular initial condition, we can find the
corresponding moment of $f_1$ using \eqref{mainb} for {\it any} driving
term $\bf S$.

The proof that $h$ is given by \eqref{adjointb} is most easily carried
out by assuming that \eqref{adjointb} holds and then by proving that $h$ is
related to the general Green's function $g$ by \eqref{greena}.
Consider the equation for $g({\bf u},\tau';{\bf u}')$,
	$$\biggl({\partial\over\partial\tau'}+D\biggr)
g({\bf u},\tau';{\bf u}')=0.$$
	We multiply this equation by $h({\bf u},\tau-\tau')$, integrate
over all velocity space, and use the adjoint relation between $D$ and
$D^*$ to give
	$$\int\dvu\,
h({\bf u},\tau-\tau'){\partial\over\partial\tau'}g({\bf u},\tau';{\bf u}')
+g({\bf u},\tau';{\bf u}')D^*\bigl(h({\bf u},\tau-\tau')\bigr)=0.$$
	Substituting from \eqref{adjointb} and integrating in $\tau'$
from $0$ to $\tau$, gives
	$$\int\dvu\,
h({\bf u},\tau-\tau')g({\bf u},\tau';{\bf u}')\biggr|_{\tau'=0}^\tau=0.$$
	If we evaluate this expression using the initial conditions for
$g$ and $h$, we obtain \eqref{greena}.

Up until now, we have assumed that all the equations are solved in an
infinite velocity domain.  This is not a convenient formulation for
numerical implementation where, necessarily, we wish to solve
equations in a finite domain.  Here we shall only solve
\eqref{adjointb} in a spherical domain $V$ such that $u<u_b$.  We will
choose $u_b$ to be sufficiently large that the interesting physics where
the electric field competes with the collisions happens inside $V$.
Outside $V$ collisions may be ignored and the electrons are merely
freely accelerated by the electric field.  We must impose boundary
conditions on $\Sigma$, the boundary of $V$.  Again we follow the
treatment given in Ref.~\citenum{Fischa}.  We begin by noting that both
\eqref{linb} and \eqref{adjointb} are hyperbolic in the $u$ direction.
The boundary therefore divides into two pieces depending on whether the
characteristics enter or leave the domain.  We define $\sigin$
(resp.~$\sigout$) as that portion of $\Sigma$ on which $-\mu-1/v^2<0$
(resp.~$\null>0$).  The characteristics of \eqref{linb} enter $V$ on
$\sigin$ and leave on $\sigout$, while those of \eqref{adjointb} enter
$V$ on $\sigout$ and leave on $\sigin$.  Boundary conditions must be
specified where the characteristics enter the domain $V$.  For $u_b$
sufficiently large, the solution beyond $\Sigma$ may be determined by
ignoring collisions.  Thus
	$$f_1({\bf u},\tau)=f_1({\bf u}+\tau{\bf \hat u}_\parallel,0)=0$$
	for $\bf u$ on $\sigin$ and
	$$h({\bf u},\tau)=h({\bf u}-\tau{\bf \hat
u}_\parallel,0)= h_0({\bf u}-\tau{\bf \hat u}_\parallel)$$
	for $\bf u$ on $\sigout$.  (If energy scattering had been
included in the collision operator, the equations would revert to
parabolic, and boundary conditions would have to be specified over the
whole of $\Sigma$.  However, there would be a boundary layer where the
characteristics of the approximate equations are outgoing, and the
boundary conditions here would only weakly affect the solution in the
interior of $V$.)

Although \eqref{adjointb} was derived under the simplifying assumptions
that the electric field was constant and the high-velocity form of
collision operator is valid, the adjoint method as described in
\refref{Fischa} applies equally well without such restrictive
assumptions.  Thus, the equation adjoint to \eqref{lin} reads
\cite{Fischa}
$$
\biggl({\partial\over\partial t'}
- {q{\bf E}(t-t')\over m}
\cdot{\partial\over \partial \bf v}-C^*\biggr) h({\bf v},t';t) =
q_1+q_2{\cal E},\Eqlab{adjoint}
$$
where $C^\ast$ is the operator adjoint to $C$.  Since the full
linearized collision operator is self-adjoint, we have $C^*(h)=C(f_mh)/f_m$.
\Eqref{adjoint} is to be solved with
the initial condition
$h({\bf v},t'=0;t)=h_0({\bf v})$.  We restrict $f_mh$ to being
orthogonal to $1$ and ${\cal E}$ and
$q_1$ and $q_2$ are chosen to ensure
that this condition on $f_mh$ remains satisified given that it is
satisfied initially.  It can then be shown \cite{Fischa} that
	$$\int\!\dvv\, f_1({\bf v},t)h_0({\bf v})
=\int_0^t\! dt'\int\!\dvv\,{\bf S}_*({\bf v},t')\cdot
{\partial\over\partial\bf v}h({\bf v},t-t';t),\Eqlab{main}$$
	where ${\bf S}_*({\bf v},t)={\bf S}({\bf v},t)
+\bigl(q{\bf E}(t)/m\bigr)f_m$.  This equation will enable us to
incorporate the effects of a slowly varying electric field into our
analysis.  It also makes explicit the additive nature of those effects
due to the electric field alone (i.e., with
${\bf S}_*=(q{\bf E}/m)f_m$) and those
effects due to the rf combined with the electric field (i.e., with
${\bf S_*}={\bf S}$). Of course, the effects due to the electric field
alone are well studied and give rise to phenomena such as the
Spitzer--H\"arm conductivity \cite{Spitzer} and runaways
\cite{Dreicer}.

\section{SOLUTIONS TO THE ADJOINT EQUATION}\label{results}

Moments of the electron distribution $f_1$ can now be calculated by
solving \eqref{adjointb} with the corresponding initial and boundary
conditions.  In practice, this procedure still offers us too much
information.  Both for a deeper understanding of the underlying physics
and for easy implementation in numerical codes, the trick is to
discover the few important functions by which the major effects can be
described.  In this section, we determine those functions needed for an
accurate treatment of rf current ramp-up.

Let us suppose that rf flux is present only for some finite time.
Electrons obeying \eqref{linb} then eventually suffer one of two fates.
Either they run away under the influence of the electric field
$u_\parallel\rightarrow-\infty$, or else they collapse into the
electron bulk $u\rightarrow 0$.  We classify these two groups of
electrons as ``runaway'' (subscript $r$) and ``stopped'' (subscript
$s$) respectively.  In a real plasma, i.e., $T\ne0$, even the bulk
particles will eventually run away.  However the time $t_r$ it takes for
these bulk electrons to run away is exponentially large, i.e., $\log
t_r\sim (v_r/v_t)^2$.  Our analysis is valid for times short compared
with the bulk runaway time.

	Runaways are very important in the calculation of the ramp-up
efficiency because runaways gain energy at the expense of the poloidal
magnetic field.  Unless they are lost, even a small number of runaways
can greatly reduce the ramp-up efficiency.  Runaways may be defined as those
particles with $u>u_0$ for $\tau\rightarrow\infty$ where $u_0$ is some
arbitrary positive speed.  (The number of runaways is independent of
$u_0$.)  Therefore their number is given by \eqref{mainb} with $\tau
\rightarrow \infty$ and
$h_0({\bf u})=1$ for $u>u_0$ and $0$ otherwise.  The Green's function
for the runaway number is given by $R({\bf u})\equiv h({\bf
u},\tau\rightarrow\infty)$ where $R$ obeys
	$$D^*\bigl(R({\bf u})\bigr)=0,\Eqlab{runaway}$$
	with boundary condition $R({\bf u})=1$ on $\sigout$.  This
function is the ``runaway probability,'' the probability that an
electron initially at $\bf u$ runs away under the combined influence of
the electric field and collisions.

	\Eqref{runaway} was solved numerically with the boundary at
$u_b=10$.  A term $\partial R/\partial\tau$ was included on the left-hand
side, and the resulting equation was integrated until $\tau=100$.  A
spherical $(u, \theta=\arccos\mu)$ grid was used with a mesh size of $500
\times100$.  The equation was integrated with an {\sc adi} (alternating
direction implicit) scheme with a time step $\Delta\tau=0.01$.  The
same method was used to solve the other equations given below.

	In \figref{run-fig}, we plot $R({\bf u})$ for $Z=1$.  For $u<1$,
$R$ is identically zero because the magnitude of the electrical force
is less than that of the frictional force.  One of the most important
applications of these results is to drive current by lower-hybrid
waves.  In this case $\bf S$ is in the parallel direction and is
localized near $u_\perp=0$.  Therefore, we need only know
$R(u_\parallel,u_\perp=0)$ which is plotted in \figref{run-figa} for
$Z=1$, 2, 5, and 10.  From this plot, we see that we can effectively
avoid the creation of runaways by operating with waves whose phase
velocities lie in the range $0<u_\parallel\lsapprox 1.5$.

	The next important quantity to determine is the current density
carried by $f_1$.  This is given (in units of $qnv_r$) by \eqref{mainb}
with $h_0({\bf u})=u_\parallel$.  The Green's function for the current
$j({\bf u},\tau)$ is therefore given by $\partial j/\partial\tau +
D^*(j)=0$ with initial condition $j(\tau=0)=u_\parallel$ and boundary
condition $j=u_\parallel-\tau$ on $\sigout$.  This is the mean current
(in units of $qv_r$) carried by an electron initially at velocity $\bf
u$.  In \figref{jt}(a), we plot $j({\bf u},\tau)$ as a function of
$\tau$ for ${\bf u}=5{\bf\hat u}_\parallel$ and $Z=1$.  Because the
presence of runaways leads to a secular behavior ($j\sim \tau$) for
large times, it is helpful to distinguish the current carried by
stopped and runaway electrons.  We write
	$$j({\bf u},\tau)=\bigl(1-R({\bf u})\bigr)j_s({\bf u},\tau)
		+R({\bf u})j_r({\bf u},\tau).$$
	The quantity $j_s$ (resp.~$j_r$) is the mean current carried by
an electron given that it eventually stops (resp.~runs away).  An
electron at velocity $\bf u$ runs away with probability $R({\bf u})$.
Thus it contributes $\bigl(1-R({\bf u})\bigr)u_\parallel$ to the
stopped current and $R({\bf u})u_\parallel$ to the runaway current.
These quantities are therefore the initial conditions to the adjoint
equations for $(1-R)j_s$ and $Rj_r$ respectively, so that
	$$\biggl({\partial\over\partial\tau}+D^*\biggr)(1-R)j_s=
\biggl({\partial\over\partial\tau}+D^*\biggr)Rj_r=0,$$
	with $j_s(\tau=0)=j_r(\tau=0)=u_\parallel$ and
$j_s=j_r=u_\parallel-\tau$ on $\sigout$.

The stopped and runaway currents $j_s$ and $j_r$ are plotted in
Figs.~\ref{jt}(b) and (c) for same case as \figref{jt}(a).  Evidently, $j_s$
vanishes for $\tau\rightarrow\infty$ (since the electrons cease to
carry any current once they are stopped).  The time it takes for the
electrons to be stopped is of the order of $u$.  Assuming that this time
is short compared to the time scale for the variation of the rf flux
$\bf S$, we may replace $j_s({\bf u},\tau)$ by $W\!_s({\bf
u})\delta(\tau)$ where $W\!_s({\bf u})=\int_0^\infty j_s({\bf
u},\tau)\,d\tau$.  The equation for $W\!_s$ is obtained by integrating
\eqref{adjointb} over time to give
	$$D^*\bigl[\bigl(1-R({\bf u})\bigr)W\!_s({\bf u})\bigr]=
		\bigl(1-R({\bf u})\bigr)u_\parallel,\Eqlab{Ws-eq}$$
	with $(1-R)W\!_s=0$ on $\sigout$.  $W\!_s$ can be interpreted as the
energy (in units of $mv_r^2$) imparted to the electric field by an
electron as it slows down. 
In \figref{W-fig}(a), we plot $W\!_s({\bf u})$ for $Z=1$. In the limit
$u_\parallel\rightarrow\infty$, collisions are extremely weak, and all
of the kinetic energy of the stopped particles goes into the electric
field, i.e.,
	$$W\!_s(u_\parallel\rightarrow\infty,u_\perp=0)
	\rightarrow{\textstyle {1\over2}} u_\parallel^2.$$
	In the limit $u\ll 1$, the electric field weakly
perturbs the electron motion.  Then, $W\!_s$ is given by the theory
of steady-state current drive \cite{Fisch-Boozer,Antonsen}, and corrections
linear in the electric field are given by the hot conductivity
\cite{Fisch-cond}.  In our notation, these results may be summarized by
	$$W\!_s(u\ll1,\mu)=
    	{\mu u^4\over 5+Z}-{(2+Z + 3\mu^2)u^6\over3(3+Z)(5+Z)}.
	\Eqlab{Ws-anal}$$
	 This function is plotted in \figref{W-fig}(b). (An
approximation to $W\!_s$ correct to order $u^{10}$ is given in
Appendix~\ref{fits}.)  This linearized
theory, however, is inapplicable for $u\sim1$, and the behavior of this
function is completely wrong for $u\grapprox1$.

Let us now turn to the contributions of the runaways to the
current.  The leading order contribution to $j_r$ is $-\tau$.  Let us
therefore write
	$$j_r({\bf u},\tau)=-\tau+j_{r0}({\bf u})
	+ j_r'({\bf u},\tau),\Eqlab{run-curr}$$
	where $j_r'(\tau\rightarrow\infty)\sim 1/\tau$ and
$j_{r0}$ may be interpreted as the effective starting velocity
for the runaways; see \figref{jt}(c).  The function
$j_{r0}$ is given by
	$$D^*\bigl(R({\bf u})j_{r0}({\bf u})\bigr)=R({\bf u}),$$
	with boundary condition $j_{r0}=u_\parallel$ on $\sigout$.  This
function is shown in \figref{u0-fig}.  For $u\gg1$, the runaway
electrons are only weakly perturbed by collisions so that $j_{r0}({\bf
u})\approx u_\parallel$.  Close to $u=1$, collisions hold back the
runaway electrons and $j_{r0}({\bf u})$ becomes large.
However, it is not very important to know
$j_{r0}$ and $j_r'$ very accurately since they are usually dominated by
the first term in \eqref{run-curr}.  We will approximate $j_{r0}({\bf
u})$ by $u_\parallel$ and will ignore $j_r'({\bf u},\tau)$ to give
$j_r({\bf u},\tau)=u_\parallel-\tau$.

Finally, we can write the Green's function for the current in an
expedient form as
	$$j({\bf u},\tau)=
\bigl(1-R({\bf u})\bigr)W\!_s({\bf u})\delta(\tau)+R({\bf u})
(u_\parallel-\tau).
\Eqlab{curr-b}$$
	In this form, it depends only on two scalar functions of $\bf
u$, namely $R$ and $W\!_s$.  Approximate fits to these functions are given
in Appendix~\ref{fits}.  An easy but important generalization is
possible here and that is to allow a loss mechanism for runaways.  This
is done by modifying suitably the term $u_\parallel-\tau$.  For example,
if the loss of runaways can be characterized by a loss time $\tau_{\rm
loss}$, this term should be multiplied by $\exp(-\tau/\tau_{\rm
loss})$.

Using this formulation, many other moments of $f_1$ may be found.  For
example, we may wish to know the mean perpendicular energy of the
runaway particles ${\cal E}_{\perp r}$ (in units of $mv_r^2$) as they
leave the integration region $V$.  (The loss rate for runaways may
depend on this quantity.)  This is given by
	$$D^*\bigl(R({\bf u}){\cal E}_{\perp r}({\bf u})\bigr)=0,$$
	with ${\cal E}_{\perp r}={1\over2}u_\perp^2$ on $\sigout$.
(This result depends logarithmically on the value of $u_b$.)
We have plotted this in \figref{E-fig}.  For electrons with
$u_\parallel>1$ and $u_\perp=0$, ${\cal E}_{\perp r}$ is about 3.  This
reflects the necessity for the electrons to suffer appreciable
pitch-angle scattering if they are to run away.

\section{CIRCUIT EQUATIONS}\label{circuit}

When rf energy is injected into a tokamak, it induces a flux ${\bf S}$ of
electrons in velocity space.  The power deposited per unit volume is
then given by
	$$p_{\rm rf}(t)=\int\!\dvv\, m{\bf S}({\bf v},t)
\cdot{\bf v}.\Eqlab{pd}$$
	Here, $p_{\rm rf}$, $\bf S$, and the other intensive physical
quantities introduced in this section also depend on position $\bf r$.
For brevity, this dependence is not shown in the arguments to these
quantities.  This equation may be used in two ways.  In detailed studies
of rf current ramp-up, based for instance on a ray-tracing model, we
can estimate ${\bf S}({\bf v},t)$ on each flux surface by
solving a one- or two-dimensional Fokker--Planck equation.  \Eqref{pd}
then gives us the power deposition, $p_{\rm rf}$.  Alternatively, we can
take the experimental measurements together with an energy balance of
the rf energy to give us an estimate of $p_{\rm rf}$.  This, together
with an approximate knowledge of where in velocity space the rf flux is
localized, allows us to determine ${\bf S}$.  In addition to causing
power absorption, the flux ${\bf S}$ leads to numerous other effects,
such as rf-driven currents, rf-enhanced particle transport, etc.  Here
our primary concern is with the rf-driven current.  From \eqref{main},
we see that this enters additively to the ohmic current so that the
total current density is given by the constitutive relation
	$$J(t)=\sigma(t) E(t)+J_{\rm rf}(t),\Eqlab{jtot}$$
	where $\sigma(t)$ is the Spitzer--H\"arm conductivity
\cite{Spitzer} for a Maxwellian plasma characterized by the background
electron temperature $T(t)$, and $J_{\rm rf}$ is the rf-driven current
density.  Here we have assumed that $\abs{v_r}\gg v_t$ so that in the
absence of any rf we can ignore runaways.  Incorporation of this effect
merely requires the addition of the current carried by the Dreicer
runaway electrons in \eqref{jtot}.

The rf-driven current density is given by \eqref{main} with the $h$
replaced by the current Green's function $j$ and with ${\bf S}_*={\bf
S}$.  Let us begin by writing $j$ in unnormalized units.  The form
for $j$ given in \eqref{curr-b} will be sufficiently accurate for our
purposes.  Multiplying by $qv_r$ gives
	$$j({\bf v},t)={qv_r\over \nu_r}
\bigl(1-R({\bf u})\bigr)W\!_s({\bf u})\delta(t)+qR({\bf u})
\biggl(v_\parallel+{qE\over m}t\biggr),$$
	where ${\bf u}={\bf v}/v_r$.  Here $j$ is now a dimensional
quantity, but $W\!_s$ and $R$ remain dimensionless functions of a
dimensionless argument.  In deriving this form for $j$ we assumed that
$E$ and $n$ were constant.  We now relax this constraint, allowing them
both to vary on a time scale long compared to the runaway collision time
$\nu_r^{-1}$.  (Recall that ${\bf S}$ is also allowed to vary on the
same time scale.)  We can then write $j$ as
	$$
j({\bf v},t-t';t)={qv_r(t')\over \nu_r(t')}
\bigl(1-R({\bf u}')\bigr)W\!_s({\bf u}')\delta(t-t')
+qR({\bf u}')\biggl(v_\parallel+{q\over m}\int_{t'}^t E(s)\,ds\biggr)
,\Eqlab{current}$$
	where ${\bf u}'={\bf v}/v_r(t')$.  The additional parametric
argument $t$ here has the same meaning as in \eqref{adjoint}.  In this
form $j({\bf v},t-t';t)$ is the mean current carried by an electron at
time $t$ given that it was traveling at velocity ${\bf v}$ at time
$t'$.  From \eqref{main}, the rf generated current density may now be
written as
	$$J_{\rm rf}(t)
=\int_0^t\! dt'\int\!\dvv\,{\bf S}({\bf v},t')\cdot
{\partial\over\partial\bf v}j({\bf v},t-t';t).\Eqlab{jrf}$$

In order to write $J_{\rm rf}$ in a more useful form, we first define a
runaway density $n_r$ (in electrons per unit volume).  This is given by
	$${\partial n_r(t)\over \partial t} = 
{1\over v_r(t)}
\int\!\dvv\,{\bf S}({\bf v},t)\cdot
{\partial\over\partial\bf u}R({\bf u}),\Eqlab{run}$$
	with initial condition $n_r(t=0)=0$.
Substituting \eqref{current} into \eqref{jrf}, we obtain
	$$J_{\rm rf}(t)=J_s(t)+J_r(t),\eqn(\eqlab{jrf-a}a)$$
	where
	$$\eqalignno{J_s(t)&={q\over \nu_r(t)}
\int\!\dvv\,{\bf S}({\bf v},t)\cdot
{\partial\over\partial\bf u} \bigl(1-R({\bf u})\bigr)W\!_s({\bf u}),
&(\ref{jrf-a}b)\cr
	{\partial J_r(t)\over\partial t}&=
{q^2\over m} E(t)n_r(t)+
q
\int\!\dvv\,{\bf S}({\bf v},t)\cdot
{\partial\over\partial\bf u}R({\bf u})u_\parallel,
&(\ref{jrf-a}c)\cr}$$
	with $J_r(t=0)=0$.  In Eqs.~(\ref{run}) and (\ref{jrf-a}), $\bf
u$ is normalized in terms of the runaway velocity at time $t$, ${\bf u}
= {\bf v}/v_r(t)$.  These equations allow the current to be calculated
by characterizing the runaway population with just two state variables
$n_r$ and $J_r$.  Equations~(\ref{pd}), (\ref{jtot}), (\ref{run}), and
(\ref{jrf-a}) suffice to give a detailed description of rf current
ramp-up.  In this form, \eqref{jtot} is suitable for substituting into a
transport or ray-tracing code.  Furthermore, it would be easy to modify
\eqref{run} to include a loss mechanism for the runaways.  Relativistic
effects on the runaways could be included in an approximate fashion by
limiting $\abs{J_r(t)/qn_r(t)}$ to $c$ the speed of light.  Such effects
could be treated in a more systematic manner by modifying the term in
large parentheses in \eqref{current} to read $v_\parallel(t)$ where
	$$\eqalign{v_\parallel(t)&=p_\parallel(t)/m\gamma,\cr
	p_\parallel(t)&=mv_\parallel+\int_{t'}^t qE(s) \,ds,\cr
	\gamma&=\sqrt{1+p_\parallel^2(t)/m^2c^2}.\cr}$$
The resulting expression for $j({\bf v},t-t';t)$ is valid for
$v_r^2\ll c^2$ and $v^2\ll c^2$.  Unfortunately, this is a
significantly more cumbersome expression from which to calculate
$J_{\rm rf}$ because, in order to determine
the state of the plasma at a particular instant, the entire runaway
distribution must be given (instead of just $n_r$ and $J_r$).

\section{APPLICATIONS}\label{applications}

The circuit equations written in Sec.~\ref{circuit} allow us to 
explore how $J_{\rm rf}$ interacts with the electric field to
yield an efficient conversion of rf energy into poloidal magnetic field
energy.  It is helpful to convert to extensive physical quantities by
assuming that the plasma current is carried in a channel of area $A$ in
which the plasma properties are approximately uniform.  Thus, the total
current is given by $I=AJ$, the total rf power deposited in the
electrons by $P_{\rm in}=2\pi R_0 Ap_{\rm rf}$ (where $R_0$ is the
tokamak major radius), the loop voltage by $V=2\pi R_0E$, etc.  The
plasma current is again written as the sum of ohmic and rf
contributions
	$$I=V/R_{\rm Sp}+I_{\rm rf},\Eqlab{I-eq}$$
	where $R_{\rm Sp}=2\pi R_0/A\sigma$ is the plasma
(Spitzer--H\"arm) resistance.  Faraday's law relates the rate of change
of the current to the voltage
	$$V=-L\dot I+V_{\rm ext},\Eqlab{V-eq}$$
	where $L$ is the total plasma inductance, which for simplicity
we shall take to be constant, $V_{\rm ext}$ is the voltage induced
by the external coils (usually a combination of the ohmic windings and
the vertical field coils), and $\dot I\equiv
dI/dt$.  Multiplying this equation by $I$ and
substituting for $I$ from \eqref{I-eq} gives
	$$\dot W=P_{\rm ext}+P_{\rm el}-{V^2\over R_{\rm Sp}},
\Eqlab{W-eq}$$
	where $W\equiv {1\over 2}LI^2$ is the poloidal field energy,
$P_{\rm ext}\equiv V_{\rm ext}I$ is the power coupled from the external
circuits, and $P_{\rm el}\equiv -VI_{\rm rf}$ is
the power coupled from the rf source into
electromagnetic energy.  This equation describes the energy balance for
the poloidal magnetic field.  The practical measure of the efficiency
of current ramp-up is
	$${\dot W-P_{\rm ext}\over P_{\rm rf}}=
{P_{\rm el}-{V^2/R_{\rm Sp}}\over P_{\rm rf}},\Eqlab{eff-eq}$$
	where $P_{\rm rf}$ is the total rf power injected into the
plasma.  The rf power absorbed by the electrons $P_{\rm in}$ is related
to $P_{\rm rf}$ by $P_{\rm in}=\eta P_{\rm rf}$ where $\eta$ is the
absorption factor.  The determination of $\eta$ is beyond the scope of
this paper; presumably it can be found by ray-tracing theories or by a
power balance.  The overall picture of the flow of power in an
experiment is as follows:  Rf power $P_{\rm rf}$ is injected into the
machine.  Of this a fraction $\eta$ is absorbed by the resonant
electrons; the rest may be absorbed by the ions or by the vacuum
vessel.  A fraction $P_{\rm el}/P_{\rm in}$ of this power is then
converted into electromagnetic energy.  $P_{\rm ext}$ acts as another
source of poloidal field energy, while the ohmic dissipation $V^2/R_{\rm Sp}$
acts as a drain.  From this discussion, we see that 
$P_{\rm el}/P_{\rm in}$ describes the ``ideal'' efficiency of rf
current ramp-up.  The practical efficiency is expressible in terms of
this efficiency, $\eta$, and $V^2/R_{\rm Sp}$.

The determination of $P_{\rm el}/P_{\rm in}$ from \eqref{jrf-a} is
complicated by the presence of runaways.  Runaways are deleterious to
the ramp-up efficiency since their current is in the same direction as
$E$ and so they subtract from $P_{\rm el}$.  For efficient current
ramp-up we must either avoid creating runaways by making sure $\bf S$
is localized in that region of velocity space where the runaway
probability $R$ is small (see Figs.~\ref{run-fig} and \ref{run-figa}),
or else take steps to lose the runaways.  We
can approximately treat these cases by taking $R=0$ in \eqref{jrf-a} to
give
	$${P_{\rm el}\over P_{\rm in}}
={\int\!\dvu\,{\bf S}\cdot \partial W\!_s/\partial {\bf u}\over
\int\!\dvu\,{\bf S}\cdot {\bf u}}.$$
	Since this involves the ratio of two integrals over $\bf u$,
the result is insensitive to the detailed form of $\bf S$.  In cases of
practical interest, we may assume that $\bf S$ is localized in $\bf u$.
Then, we have
	$${P_{\rm el}\over P_{\rm in}}= {{\bf\hat S}\cdot \partial
W\!_s/\partial {\bf u}\over {\bf\hat S}\cdot {\bf u}},\Eqlab{eff}$$
	where ${\bf u}$ is the normalized velocity of the resonant
electrons.

For lower-hybrid waves we have ${\bf\hat S}={\bf \hat u}_\parallel$ and
the waves interact with particles through the Landau resonance
$\omega-k_\parallel v_\parallel=0$, where $\omega$ and $k_\parallel$ are
the wave frequency and parallel wave number.  Furthermore, the typical
perpendicular velocity of the resonant electrons equals the electron
thermal velocity, so that $v_\perp\sim v_t\ll v_\parallel$.  Thus
\eqref{eff} is to be evaluated with $u_\parallel = \omega/k_\parallel
v_r$ and $u_\perp=0$.  This gives
	$${P_{\rm el}\over P_{\rm in}}= 
{\partial W\!_s/\partial u\over u}.\eqn(\eqlab{effa}a)$$
	This efficiency is plotted in \figref{W-figb}(a). 
Approximate fits for this function are given in Appendix~\ref{fits}.

On the other hand, for electron-cyclotron waves which interact through
the Doppler-shifted cyclotron resonance $\omega-k_\parallel v_\parallel
= l \Omega$, where $\Omega$ is the cyclotron frequency and $l$ is the
harmonic number, we have ${\bf\hat S}={\bf \hat u}_\perp$.  In this case
we evaluate \eqref{eff} at $u_\parallel = (\omega-l\Omega)/k_\parallel
v_r$ and $u_\perp=0$ to give
	$${P_{\rm el}\over P_{\rm in}}= 
{\partial W\!_s/\partial u-(1/u_\parallel)\partial W\!_s/\partial\mu
\over u},\eqn(\ref{effa}b)$$
	which is plotted in \figref{W-figb}(b).

Using Eqs.~(\ref{effa}), it is possible to identify regions of high
conversion efficiency of wave energy to electric energy, given the
restriction of $R$ small.  Additionally, if the ohmic losses,
$V^2/R_{\rm Sp}$, are small, then by \eqref{eff-eq}, we see that the
conversion of wave energy to poloidal field energy can be of high
efficiency.  This, in fact, is what has been achieved on the {\sc plt}
experiment, where conversion efficiencies of over $25\%$ have been
reported \cite{Jobes}.

An important practical consequence of the circuit
equations derived here is that fast ramp-up rates, i.e., large $\dot I$,
are possible at high density.  In fact, these fast ramp-up
rates are {\it necessary} for high energy conversion efficiencies at high
density.   This can be seen as follows:  The
efficiency, $P_{\rm el}/P_{\rm in}$, is a function of the dimensionless
parameter $u_\parallel$, depending, in addition, only weakly on
$Z$.  For a given machine and a given wave phase velocity, the parameter
$u_\parallel$ depends only on the ratio, $E/n$; and the ramp-up rate,
$\dot I$, depends only on the dc electric field, $E$.  Thus, the
efficiency depends only on the ratio of $\dot I$ to $n$.  It has been
observed experimentally on the {\sc PLT} experiment that high
efficiency of converting rf energy to magnetic field energy is possible
at a low plasma density.  Thus, we can predict that a similar high
efficiency is possible in the event that the density and the ramp-up
rate are scaled up together.  In fact, for large ramp-up rates, high
density can actually be desirable in that it impedes the production of
runaways.  Note that this window of desired density for a given ramp-up
rate is counter to our intuition derived from steady-state
considerations, where the larger the density the less the current-drive
efficiency.

There are several optimizations that one might wish to achieve in the
ramp-up problem.  One is to maximize the energy conversion efficiency,
$P_{\rm el}/P_{\rm in}$.  A second is to minimize the ramp-up time,
$T_{\rm ramp}\equiv I/\dot I$.  The minimization of capital costs for
the rf system, however, may demand that we minimize $P_{\rm rf}$, the
rf power required to ramp-up a given current.

We can express this more precisely with some convenient formulas.  In
the absence of the external source $V_{\rm ext}$, the
ramp-up rate may be written using Faraday's law as
	$$\dot I\approx {5E\over\ln R_0/a}\rm \,{MA\over s},\Eqlab{idot}$$
	where a tokamak inductance $L\approx \mu_0R_0\ln R_0/a$ was assumed,
and where $E$ is the dc electric field in units of $\rm V/m$.  Note
that the ramp-up rate depends linearly on $E$ and is almost
independent of geometry ($\ln R_0/a\approx 1$).  The amount of dissipated
rf power required can than be written as
	$$P_{\rm rf}\approx {1\over \eta}
{{1\over 2}LI^2\over T_{\rm ramp}}\bigg/
{P_{\rm el}\over P_{\rm in}},
\Eqlab{p-req}$$
	if we neglect both ohmic losses ($V^2/R_{\rm Sp}$) and the
external source $P_{\rm ext}$.  Thus, in
extrapolating results to larger tokamaks (higher ${1\over2}LI^2$), we
can maintain linear (in the required stored energy)
power requirements, with the same ramp-up
time and the same efficiencies, if the density scales linearly with $\dot
I$ and hence with $I$.  Here, the wave phase velocities also remain the
same, and in the event of the same temperatures, the physics of the
damping may be expected to be very similar, so that the percentage of
incident rf power that is absorbed, $\eta$, remains constant too.

For example, using \figref{W-figb}(a), we see that in {\sc plt} with
$n\approx 2\times 10^{12}\,\rm cm^{-3}$, $T\approx 1\,\rm keV$,
$\omega/k_\parallel\approx 6v_t\approx{1\over 4}c$, we find reported
ramp-up rates of $\dot I=120\rm\,kA/s$, or $E\approx24 \rm\,mV/m$.
Here $u_\parallel\approx1.4$ and ideal efficiencies of about $33\%$ at $Z=1$
may be expected with little runaway production, consistent with
experimental data.  Also, consistent with the data would be somewhat
higher $Z$, but then only if the confinement of runaways were not
perfect.

For reactor-grade tokamaks, say $LI^2\approx 400\,\rm MJ$ and $I\approx
10\,\rm MA$,  a ramp-up time which is longer than that in {\sc plt} is
desirable in order to minimize $P_{\rm rf}$ and the capital cost of the
rf system.  For a $30\rm\,s$ ramp-up time, a density of
$5\times10^{12}\,\rm cm^{-3}$ renders the ratio $\dot I/n$ as in the {\sc
plt} experiment.  Employing a similar spectrum of waves
($\omega/k_\parallel = {1\over 4}c$) in a plasma of temperature also
similar to the {\sc plt} experiment ($T\approx1\,\rm keV$) implies a
similar $\eta$ (about 0.7).  Thus, using \eqref{p-req}, we see that
$P_{\rm rf}\approx 40\,\rm MW$ would be required.

To summarize the tradeoffs here, we note that while $P_{\rm el}/P_{\rm
in}$ is minimized by considering only the ratio $\dot I/n$, the
minimization of $P_{\rm rf}$ requires the $T_{\rm ramp}$ be large.
Thus, although very quick ramp-up rates are indeed achievable at high
density, the capital costs for such a system are proportionately larger
too.  Balancing the desires for a quick ramp-up against those for low
capital costs (low $P_{\rm rf}$) points to a parameter range of
moderate density.  Efficient ramp-up is only achieved when, in addition
to the above restrictions, the temperature is moderate, since at high
temperatures, $V^2/R_{\rm Sp}$ losses, neglected in \eqref{p-req},
begin to dominate.  The regime where these ohmic losses dominate may be
identified by writing
	$${V^2\over R_{\rm Sp}}\approx {LI^2\over
T_{\rm ramp}} {L/R_{\rm Sp}\over T_{\rm ramp}}.$$
	These losses represent only small corrections when $V^2/
R_{\rm Sp}\ll P_{\rm rf}$, or, using \eqref{p-req}, when
	$${T_{\rm ramp}\over L/R_{\rm Sp}}\gg {P_{\rm el}\over P_{\rm
rf}}.$$
	For the reactor-grade example, $P_{\rm el}/P_{\rm
in}\approx{1\over 3}$, $\eta\approx 0.7$, the above inequality
requires that the ramp-up time be longer than about $1\over 4$ of the
$L/R_{\rm Sp}$ time.  This restricts the temperature to somewhat less
than $2\,\rm keV$.

Restricting the temperature during a period of intense rf injection
(perhaps $40\,\rm MW$) requires a small heat confinement time during
the start-up operation.  In the above example, this may be as small as
$30\,\rm ms$.  Poor confinement during the start-up phase may be
helpful from the standpoint of runaway buildup too.  Even a small
percentage $(\sim 1\%)$ of reverse runaways \cite{Eder} can seriously impede
ramp-up if the runaways are well confined.  If the runaways are poorly
confined, then higher percentages may be tolerated, allowing higher
ramp-up rates and, consequently, higher energy conversion efficiencies
for a given density.

We are led thus to the following typical picture of rf ramp-up for
pulsed tokamak operation.  Start-up can proceed in a low density plasma
\cite{start-up} where the rf power is also used to initiate the
plasma. Density and rf power, and the ramp-up rate, are increased
concomitantly as the plasma is brought to interesting densities
$10^{13}$--$10^{14}\,\rm cm^{-3}$.    During this phase, the
temperature is purposefully kept low, possibly through a deliberate
degradation of the confinement of both runaway and thermal electrons.
Hence the current is programmed to reach a large value prior to the
density, and both reach large values prior to the temperature.  The
final step, in which the reactor is brought to reactor-grade
temperature, occurs after the current is ramped up and as a result of
ceasing the deliberate degradation of confinement.

\section{CONCLUSIONS}\label{conclusions}

In this paper we have written down a set of circuit equations that
describe the dynamics of an rf-driven plasma.  In arriving at these
circuit equations, we systematically introduced approximations with a
goal of characterizing the driven plasma by a small number of functions
of few variables that retain the essential physics.  Greater accuracy,
possible at the price of more complex circuit equations, may be
obtained as a natural extension of the development here.  The
identification and calculation here of a minimal set of transport
functions, however, provided a suitable and manageable description for a
large class of important problems.

The calculations of the runaway function $R$ and of the energy
conversion function $W\!_{s}$ together pinpoint the preferred region
for tokamak ramp-up operation.  These functions depend only on the
dimensionless parameter $\bf u$.  The separate contributions of runaway
and stopped currents may be described using these functions of a single
variable.  The constitutive relations thus obtained are given by
Eqs.~(\ref{jrf-a}).  These equations are in a form both suitable for
implementation in a transport code and amenable to obvious modification
in the event that more complex runaway models are desired.

There are several caveats to bear in mind in using these formulas.
First, the time scale for variation of the dc electric fields has been
assumed long compared to other scales of interest, such as the particle
deceleration times.  A violation of this scale separation would affect
the normalizations through $v_r$.  Second, knowledge of the rf spectrum
is unlikely to be complete.  This knowledge is necessary to give $\bf
S$, the rf-induced flux.  Even if the incident rf energy is followed by
ray-tracing codes, it remains possible that other waves may be present.
These other waves might arise either due to asymmetries in the particle
distribution functions or due to nonlinear effects associated with the
incident spectrum.  Third, particle transport across field lines was
neglected in comparison to the effects along field lines.  The neglect
of these effects is possible for stopped electrons if they are stopped
before they reach a flux surface with significantly different conditions
($v_r$ different).  For runaway electrons, these effects are always
important in that they provide a model for the runaway loss.  As
discussed after Eqs.~(\ref{jrf-a}), such model may be included through
a natural modification of Eq.~(\ref{jrf-a}c).  In the absence of one
particularly compelling model, at present, for runaway loss we have
left the modification of  Eq.~(\ref{jrf-a}c) as an open issue.

Finally, we should note that some of the most powerful conclusions of
this paper occur in certain special cases.  It is often the case that
the rf spectrum is not only known, but also localized, which enables a
particularly simple evaluation of the conversion efficiency, as in
Eqs.~(\ref{effa}).  In the event of moderate electric fields, or
spectra localized at moderate phase velocities, it may be that $R=0$
(no runaway production), and an accurate runaway loss model would not be
needed.  In the event that runaways are confined well, the spectrum
must be chosen carefully to assure that $R=0$.

\section*{ACKNOWLEDGMENTS}

This work was supported by the United States Department of Energy under
Contract DE--AC02--76--CHO--3073.

\appendix

\section{LANGEVIN EQUATIONS}\label{fokker}

Here we show that the conditional probability distribution $g({\bf
u},\tau;{\bf u}')$ for Eqs.~(\ref{monte}) satisfies \eqref{green}.  The
derivation follows those given in Refs.~\citenum{Wang} and
\citenum{Lifshitz}.  Because the process described by Eqs.~(\ref{monte})
is a Markoff process, $g$ satisfies the Smolucowski equation
\cite{Wang}
	$$
g({\bf u},\tau+\Delta\tau;{\bf u}')=\int\dvu''\,
g({\bf u},\Delta\tau;{\bf u}'')g({\bf u}'',\tau;{\bf u}')
\Eqlab{Smol}$$
	for all $\tau>0$ and $\Delta\tau>0$.
Let us define
	$$r({\bf w},{\bf u},\Delta\tau)\equiv
g({\bf u}+{\bf w},\Delta\tau;{\bf u}).$$
	Subtracting $g({\bf u},\tau;{\bf u}')$ from \eqref{Smol} gives
$$\eqalignno{
g({\bf u},\tau+\Delta\tau;{\bf u}')-g({\bf u},\tau;{\bf u}')
&=\int\dvu''\,
\bigl[r({\bf u}-{\bf u}'',{\bf u}'',\Delta\tau)g({\bf u}'',\tau;{\bf u}')\cr
&\qquad\qquad{}-r({\bf u}-{\bf u}'',{\bf u},\Delta\tau)g({\bf u},\tau;{\bf
u}')\bigr],&(\eqlab{fp1})\cr}$$
	where, because of the normalization condition for probabilities,
the second term in the integral may be reduced to $g({\bf u},\tau;{\bf
u}')$.  If we change the variable of integration to ${\bf w}={\bf
u}-{\bf u}''$, the right-hand side of \eqref{fp1} becomes
	$$\int d^3{\bf w}\,
\bigl[r({\bf w},{\bf u}-{\bf w},\Delta\tau)g({\bf u}-{\bf w},\tau;{\bf u}')
-r({\bf w},{\bf u},\Delta\tau)g({\bf u},\tau;{\bf u}')\bigr].$$
	For small $\Delta\tau$, the function $r({\bf w},{\bf
u},\Delta\tau)$ is highly localized about ${\bf w}=0$.  We may
therefore expand the first term in the integral, assuming that ${\bf w}$
is much smaller than ${\bf u}$, to give
	$$\eqalign{
r({\bf w},{\bf u}-{\bf w},\Delta\tau)g({\bf u}-{\bf w},\tau;{\bf u}')
&\approx
r({\bf w},{\bf u},\Delta\tau)g({\bf u},\tau;{\bf u}')
-{\bf w}\cdot{\partial\over\partial {\bf u}}
r({\bf w},{\bf u},\Delta\tau)g({\bf u},\tau;{\bf u}')\cr
&\qquad\qquad\qquad\qquad{}+{1\over 2}{\bf w}{\bf w}\mathbin{:}
{\partial^2\over\partial {\bf u}\partial {\bf u}}
r({\bf w},{\bf u},\Delta\tau)g({\bf u},\tau;{\bf u}').\cr}$$
	Using this approximation in \eqref{fp1}, integrating by parts,
dividing by $\Delta \tau$ and taking the limit
$\Delta\tau\rightarrow0$, we find
as the equation for $g({\bf u},\tau;{\bf u}')$
	$${\partial\over\partial\tau}g
={\partial\over\partial{\bf u}}\cdot
{\bf A}g+
{\partial^2\over\partial{\bf u}\partial{\bf u}}\mathbin{:}
\mat B g,\Eqlab{fp2}$$
	where
	$$\eqalign{
{\bf A}({\bf u})&=
\lim_{\Delta\tau\rightarrow 0}-{\ave{\Delta{\bf u}}\over\Delta \tau}
,\cr
\mat B({\bf u})&=
\lim_{\Delta\tau\rightarrow 0}{\ave{\Delta{\bf u}\Delta{\bf u}}
\over2\Delta \tau}
,\cr}$$
and
$$\eqalign{
\ave{\Delta{\bf u}}&=\int {\bf w}\,
r({\bf w},{\bf u},\Delta\tau)\,d^3{\bf w},\cr
\ave{\Delta{\bf u}\Delta{\bf u}}&=\int{\bf w}{\bf w}\,
r({\bf w},{\bf u},\Delta\tau)\,d^3{\bf w}.\cr}$$
	Thus $\ave{\Delta{\bf u}}$ is the average value of ${\bf
u}(\tau+\Delta\tau)-{\bf u}(\tau)$ given that ${\bf u}(\tau)={\bf u}$
(and similarly for $\ave{\Delta{\bf u}\Delta{\bf u}}$).  These
quantities may be calculated directly from Eqs.~(\ref{monte}) assuming
that $\Delta\tau$ is sufficiently small that $\bf u$ does not change
appreciably.  We then obtain
$$\eqalign{
\ave{\Delta u}&=-\int_\tau^{\tau+\Delta\tau}
\ave{{1\over u(\tau')^2}+\mu(\tau')}\,d\tau'\cr
&\approx-\biggl({1\over u^2}+\mu\biggr)\Delta\tau,\cr
\ave{\Delta \mu}&=\int_\tau^{\tau+\Delta\tau}
\ave{A(\tau')-
{1-\mu(\tau')^2\over u(\tau')}}\,d\tau'\cr
&\approx-\biggl({1+Z\over u^3}\mu+{1-\mu^2\over
u}\biggr)\Delta\tau,\cr
\ave{\Delta\mu\Delta\mu}&=\int_\tau^{\tau+\Delta\tau}\!\!\!
\int_\tau^{\tau+\Delta\tau}\ave{A(\tau')A(\tau'')}\,d\tau'\,d\tau''
+O(\Delta\tau^2)\cr
&\approx {1+Z\over u^3}(1-\mu^2)\Delta\tau,\cr
\ave{\Delta u\Delta u}&=\ave{\Delta u\Delta \mu}=\ave{\Delta \mu\Delta u}
=O(\Delta\tau^2).\cr}$$
	Here we have made use of the properties of $A$ given in
Eqs.~(\ref{aprop}).
Writing \eqref{fp2} in spherical coordinates and substituting
for the non-zero components of $\bf A$ and $\mat B$, we obtain
	$$\eqalignno{{\partial\over\partial\tau}g
&={1\over u^2}{\partial\over\partial u}u^2
A_ug+{\partial\over\partial\mu}A_\mu g+
{\partial^2\over\partial\mu^2}
B_{\mu\mu} g\cr
&={\partial\over\partial u_\parallel}g+{1\over u^2}
{\partial\over\partial u}g
+{1+Z\over 2u^3}{\partial\over\partial\mu}(1-\mu^2)
{\partial\over\partial\mu}g.&(\eqlab{fp3})\cr}$$
	This is the same equation as \eqref{green}.  Furthermore, from
the definition of $g$ as a conditional probability, the initial
condition for \eqref{fp3} is also the same as for \eqref{green},
namely, $g({\bf u},\tau;{\bf u}')=\delta({\bf u}-{\bf u}')$.

\section{NUMERICAL FITS}\label{fits}

In this appendix, we give approximations for some of the important
functions we have calculated.  These are suitable for incorporating into
modeling codes.  The approximations were found by choosing a suitable
analytic form containing several undetermined coefficients and
adjusting those coefficients in order to minimize the maximum relative
error.  The technique for carrying out this procedure is described in
Hastings' classic work \cite{Hastings}.  The fits were made to the
numerical data presented in Sec.~\ref{results}.  This data itself
contains errors due to the numerical methods used.  The main source of
error is due to the finite size of the numerical mesh and it is
estimated that this introduces errors on the order of a percent.
However, near $u=0$, the relative error in the numerical data for $W\!_s$
and its derivative becomes large because $W\!_s=O(u^4)$.  Thus for
$u<0.5$, the fits were made using the following analytical
approximation instead of the numerical data:
	$$\eqalign{W\!_s&={\mu u^4\over Z+5}
-{(2+Z+3 \mu^2) u^6\over 3 (3+Z) (5+Z)}
+{2 \bigl((24+19 Z+3 Z^2) \mu+(9+Z) \mu^3\bigr) u^8
\over (3+Z) (5+Z) (7+3 Z) (9+Z)}\cr
&\quad{}-{
\bigl({1041+1864 Z+1189 Z^2+316 Z^3+30 Z^4\hfill\atop\quad
{}+10 (417+497 Z+181 Z^2+21 Z^3) \mu^2
+5 (9+Z) (13+3 Z) \mu^4}\bigr) u^{10}
\over 5 (2+Z) (3+Z) (5+Z) (7+3 Z) (9+Z) (13+3 Z)}.\cr}
$$
	This result was obtained by solving \eqref{Ws-eq} for small $u$
using {\sc macsyma} \cite{MACSYMA}.  (The first two terms in this
expansion are those derived by Fisch \cite{Fisch-cond}.)

For each of the functions approximated, we give the analytic form of
the approximation, the range in which it is valid, a table of
coefficients, and the maximum relative error.  The approximations
should not be used outside the range given.  Also, note that the
relative error quoted is the error in fitting the approximation to
the numerical data which itself is in error by about a percent.

For $\mu=1$ and $1.4<u<8$, the runaway probability $R$ is approximated
by
	$$R(u,\mu=1)=\exp\Biggl({\sum_{i=0}^3 a_i(u-1)^i\over
\sum_{i=1}^3 b_i(u-1)^i}\Biggr),$$
	where $b_1=1$ and the other coefficients $a_i$ and $b_i$ are
given in \tabref{run-tab}.  The maximum relative error is $1\%$.  For
$\mu=1$ and $1<u<1.4$, the same approximation may be used with small
absolute error but large relative error.  For $u<1$ and all $\mu$ we
have $R=0$ identically.

For $\mu=1$ and $0<u<5$, the energy imparted to the electric field $W\!_s$
by stopped electrons is approximated by
	$$W\!_s(u,\mu=1)={\sum_{i=2}^4 a_iu^{2i}\over
\sum_{i=0}^3 b_iu^{2i}},$$
	where $b_0=1$ and the other coefficients $a_i$ and $b_i$ are
given in \tabref{ws-tab1}.  The maximium relative error is $2\%$.  For
$\mu=-1$ and $0<u<1$, $W\!_s$ is approximated by
	$$W\!_s(u,\mu=-1)=\sum_{i=2}^5 a_iu^{2i},$$
	where the coefficients $a_i$ are
given in \tabref{ws-tab2} and the maximium relative error is $1.5\%$.

For $\mu=1$ and $0<u<5$, the function $(\partial W\!_s/\partial u)/u$ is
is approximated by
	$${1\over u}{\partial\over\partial u}W\!_s(u,\mu=1)
={\sum_{i=1}^3 a_iu^{2i}\over
\sum_{i=0}^3 b_iu^{2i}},$$
	where $b_0=1$ and the other coefficients $a_i$ and $b_i$ are
given in \tabref{eff-tab1}.  The maximium relative error is $5\%$.  For
$\mu=-1$ and $0<u<1$, $(\partial W\!_s/\partial u)/u$ is approximated by
	$${1\over u}{\partial\over\partial u}W\!_s(u,\mu=-1)
=\sum_{i=1}^4 a_iu^{2i},$$
	where the coefficients $a_i$ are
given in \tabref{eff-tab2} and the maximium relative error is $3\%$.

\begin{thetables}{99}
\newdimen\digitwidth\setbox0=\hbox{\rm0}\digitwidth=\wd0
\newdimen\minuswidth\setbox0=\hbox{\mathsurround=0pt$-$}\minuswidth=\wd0
{\catcode`?=\active\catcode`+=\active
 \gdef\spdef{\offinterlineskip
 \catcode`?=\active\def?{\kern\digitwidth}%
 \catcode`+=\active\def+{\mathbin{\hbox to\minuswidth{\hfil}}}}}
\tableitem{run-tab}  Coefficients for approximation to $R(u,\mu=1)$.
$$\vcenter{\tabskip.5em\spdef
\halign{\strut\hfil$# $\hfil\tabskip1em&\hfil$# $\hfil
&\hfil$# $\hfil&\hfil$# $\hfil&\hfil$# $\hfil&\hfil$# $\hfil&\hfil$# $\hfil
\tabskip.5em\cr\noalign{\hrule\vskip1.5pt\hrule}
 Z& +a_0    &?+a_1    & +a_2    & +a_3    & +b_2    & +b_3    \cr\noalign{\hrule}
 1& -3.68063&?+4.23913& -4.55894& -0.39755& -1.22774& +1.41450\cr
 2& -4.97636&-16.09015& +0.83188& +0.21737& +6.84615& -0.98649\cr
 5& -4.27687&?-4.33629& +0.30338& +0.05697& +3.21315& -0.47749\cr
10& -4.94597&?-1.53482& +0.10112& +0.03087& +2.45288& -0.36896\cr
\noalign{\hrule\vskip1.5pt\hrule}}}$$
\tableitem{ws-tab1}  Coefficients for approximation to $W\!_s(u,\mu=1)$.
$$\vcenter{\tabskip.5em\spdef
\halign{\strut\hfil$# $\hfil\tabskip1em&\hfil$# $\hfil
&\hfil$# $\hfil&\hfil$# $\hfil&\hfil$# $\hfil&\hfil$# $\hfil&\hfil$# $\hfil
\tabskip.5em\cr\noalign{\hrule\vskip1.5pt\hrule}
 Z& +a_2    & +a_3    & +a_4    & +b_1    & +b_2    & +b_3    \cr\noalign{\hrule}
 1& +0.16612& -0.01495& +0.00775& +0.37136& +0.02240& +0.01645\cr
 2& +0.14200& -0.04048& +0.01145& +0.12253& +0.00384& +0.02440\cr
 5& +0.09880& -0.05152& +0.01113& -0.19484& +0.00559& +0.02362\cr
10& +0.06537& -0.03895& +0.00738& -0.32456& +0.02797& +0.01526\cr
\noalign{\hrule\vskip1.5pt\hrule}}}$$
\tableitem{ws-tab2}  Coefficients for approximation to $W\!_s(u,\mu=-1)$.
$$\vcenter{\tabskip.5em\spdef
\halign{\strut\hfil$# $\hfil\tabskip1em&\hfil$# $\hfil
&\hfil$# $\hfil&\hfil$# $\hfil&\hfil$# $\hfil
\tabskip.5em\cr\noalign{\hrule\vskip1.5pt\hrule}
 Z& +a_2    & +a_3    & +a_4    & +a_5    \cr\noalign{\hrule}
 1& -0.16483& -0.13420& +0.15346& -0.24314\cr
 2& -0.14186& -0.09297& +0.06661& -0.12870\cr
 5& -0.09975& -0.04781& +0.00606& -0.03545\cr
10& -0.06651& -0.02797& -0.00247& -0.00934\cr
\noalign{\hrule\vskip1.5pt\hrule}}}$$
\tableitem{eff-tab1}  Coefficients for approximation to
$(\partial W\!_s/\partial u)/u$ for $\mu=1$.
$$\vcenter{\tabskip.5em\spdef
\halign{\strut\hfil$# $\hfil\tabskip1em&\hfil$# $\hfil
&\hfil$# $\hfil&\hfil$# $\hfil&\hfil$# $\hfil&\hfil$# $\hfil&\hfil$# $\hfil
\tabskip.5em\cr\noalign{\hrule\vskip1.5pt\hrule}
 Z& +a_1    & +a_2    & +a_3    & +b_1    & +b_2    & +b_3    \cr\noalign{\hrule}
 1& +0.66445& -0.36032& +0.07328& +0.17769& -0.25452& +0.07278\cr
 2& +0.56760& -0.38984& +0.08634& -0.04019& -0.24673& +0.08508\cr
 5& +0.39906& -0.32879& +0.07670& -0.28281& -0.16275& +0.07436\cr
10& +0.27028& -0.23261& +0.05272& -0.39140& -0.07526& +0.04981\cr
\noalign{\hrule\vskip1.5pt\hrule}}}$$
\tableitem{eff-tab2}  Coefficients for approximation to
$(\partial W\!_s/\partial u)/u$ for $\mu=-1$.
$$\vcenter{\tabskip.5em\spdef
\halign{\strut\hfil$# $\hfil\tabskip1em&\hfil$# $\hfil
&\hfil$# $\hfil&\hfil$# $\hfil&\hfil$# $\hfil
\tabskip.5em\cr\noalign{\hrule\vskip1.5pt\hrule}
 Z& +a_1    & +a_2    & +a_3    & +a_4    \cr\noalign{\hrule}
 1& -0.63673& -1.39960& +3.37662& -4.23684\cr
 2& -0.55777& -0.80763& +1.43144& -2.03866\cr
 5& -0.39704& -0.33811& +0.23607& -0.51011\cr
10& -0.26600& -0.17342& +0.01896& -0.13349\cr
\noalign{\hrule\vskip1.5pt\hrule}}}$$
\end{thetables}
\begin{thefigures}{99}

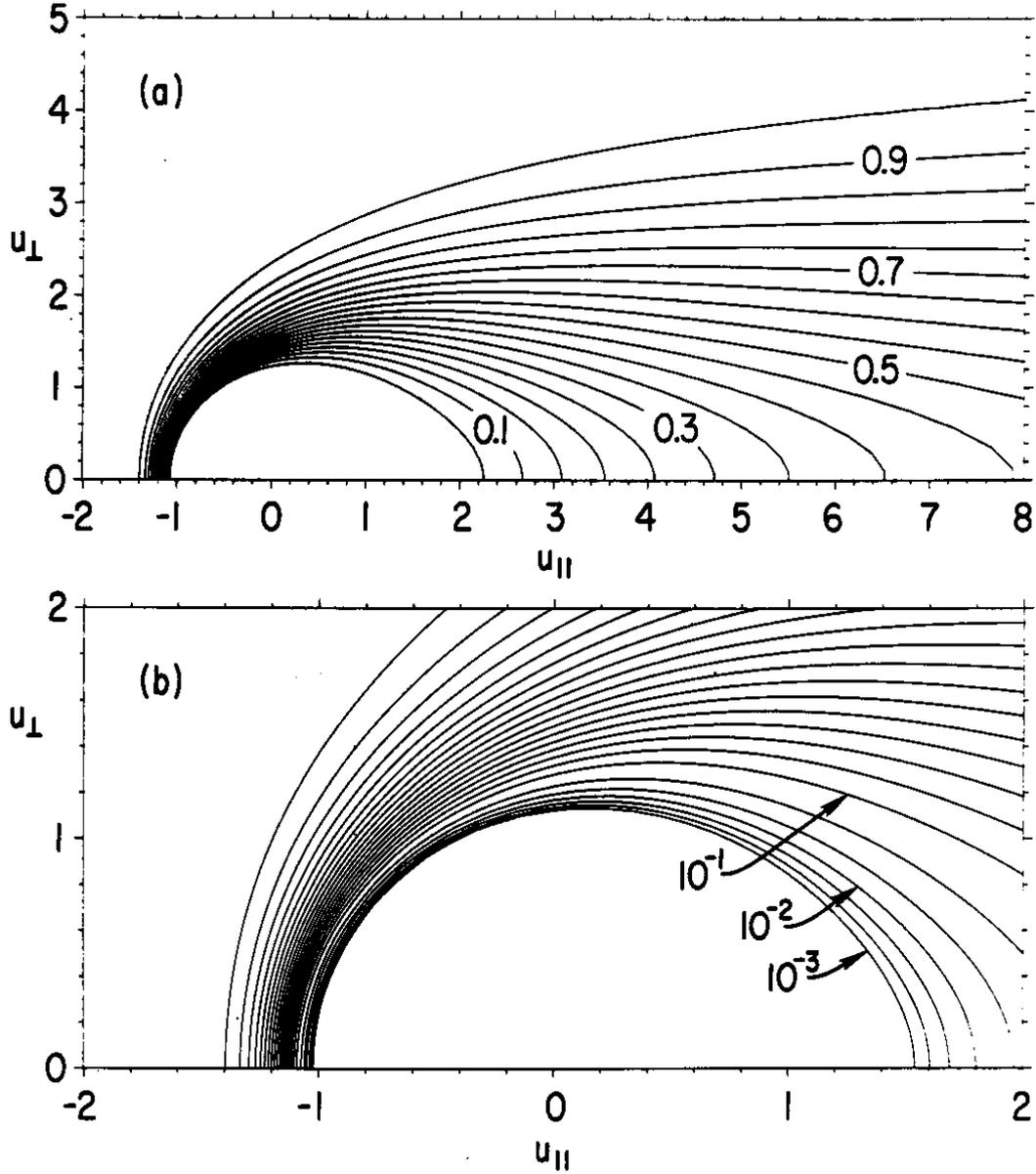
\figitem{run-fig}{5.5in} The runaway probability $R({\bf u})$ for $Z=1$.  Parts
(a) and (b) show $R$ on two different scales.  In (a) the contours are
equally spaced at intervals of 0.05.  In (b) the lowest 7 contours are
geometrically spaced at intervals of $10^{1/3}$ between $10^{-3}$ and
$10^{-1}$; the remaining contours are equally spaced at intervals of
0.05 as in (a).

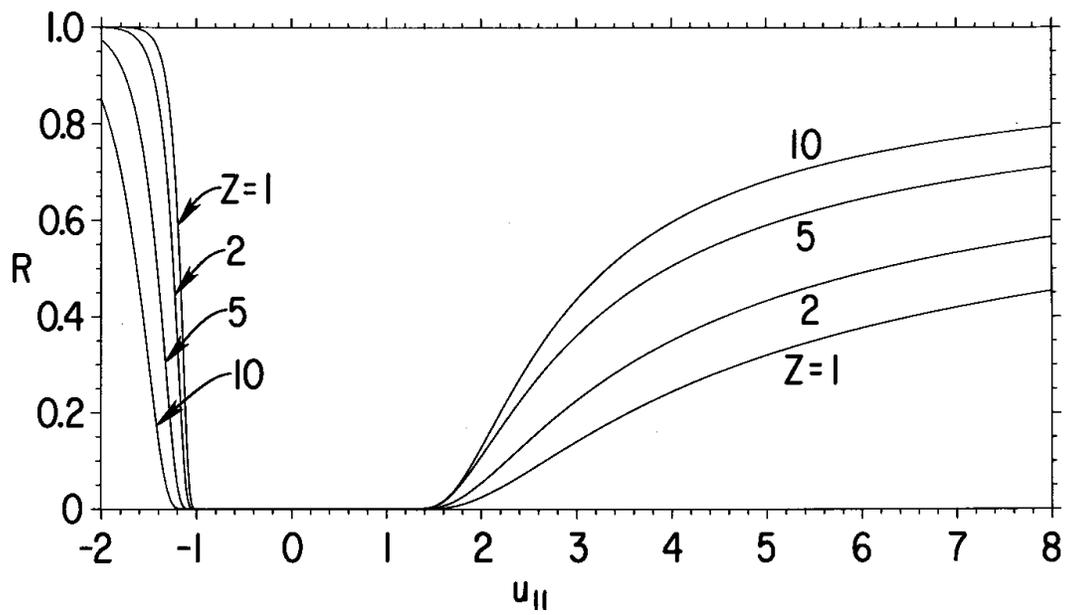
\figitem{run-figa}{5.5in} $R(u_\parallel,u_\perp=0)$ for $Z=1$, 2, 5, and 10.

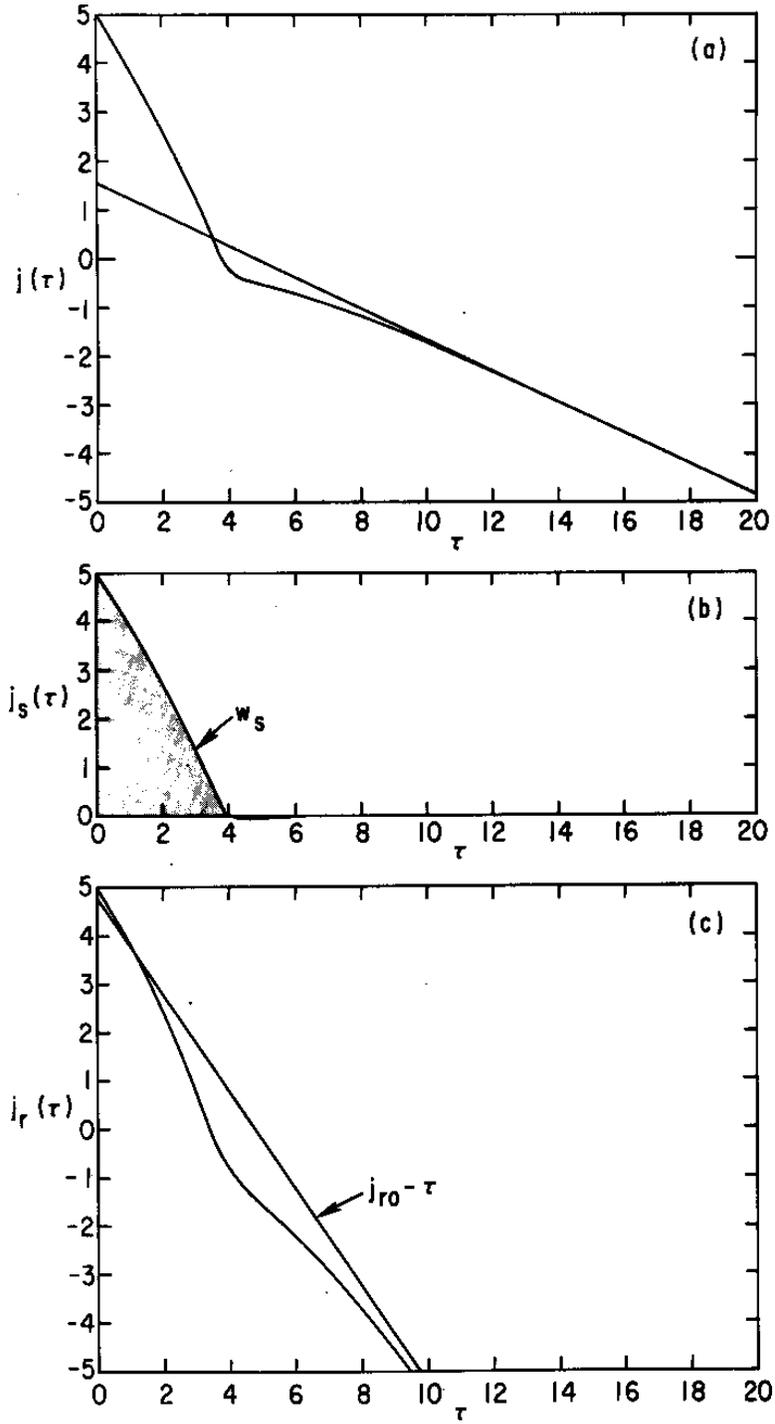
\figitem{jt}{4in} The current $j({\bf u},\tau)$ for ${\bf u}=5{\bf \hat
u}_\parallel$ and $Z=1$;  (a) the total current $j$; (b) the stopped
current $j_s$; (c) the runaway current $j_r$.  For ${\bf u}=5{\bf \hat
u}_\parallel$, approximately $32\%$ of the electrons run away.

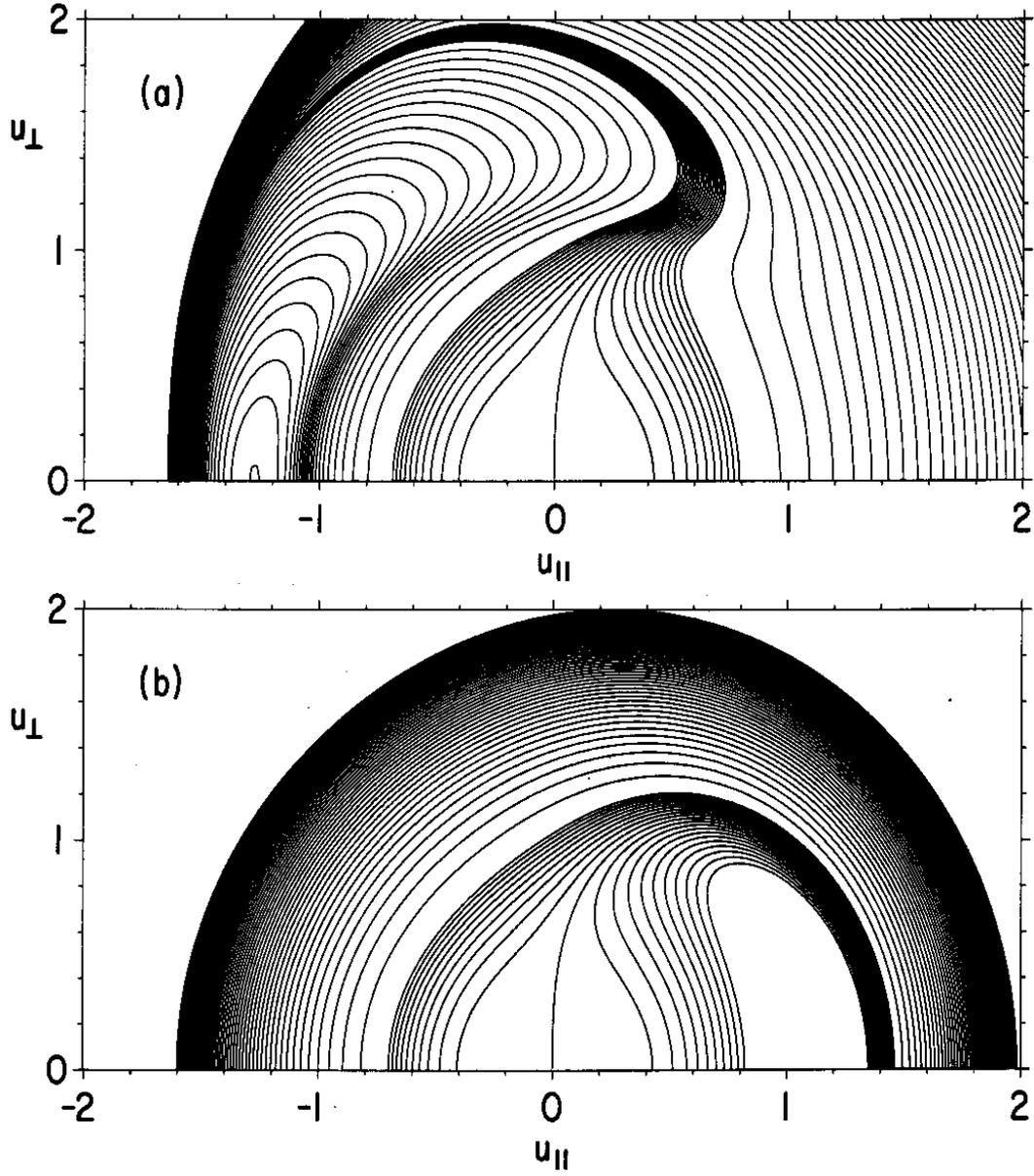
\figitem{W-fig}{5.5in} The energy imparted to the electric field by the
stopped particles $W\!_s({\bf u})$ for $Z=1$.  The innermost contours are
equally spaced at intervals of 0.005 between $-0.05$ and $0.05$.  The
remaining contours are equally spaced at intervals of 0.05.  Part (a)
shows the results of numerically solving \eqref{Ws-eq}; part (b) shows
$W\!_s$ from the hot-conductivity theory \eqref{Ws-anal}.

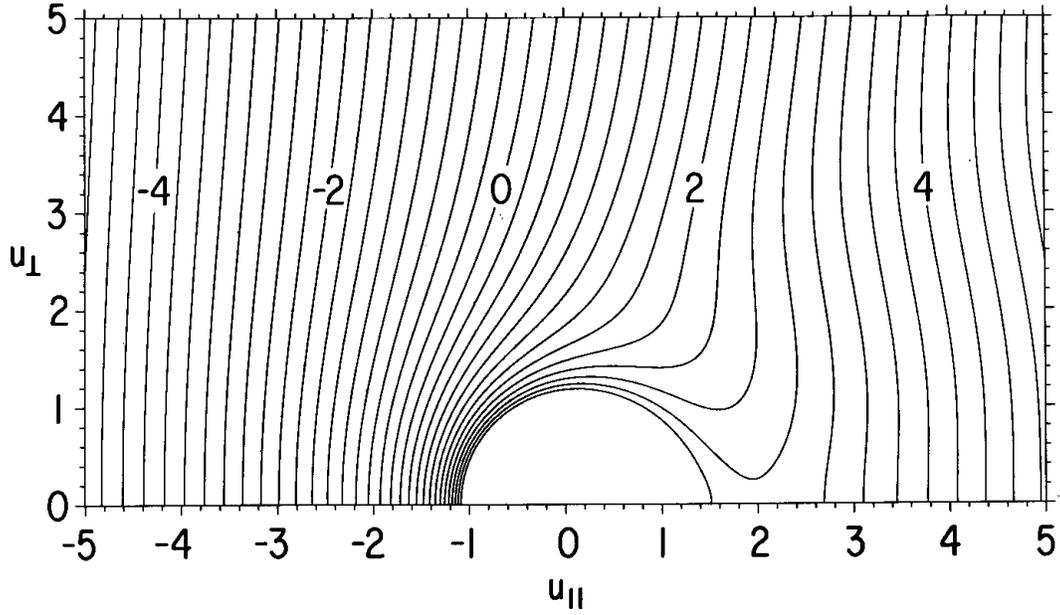
\figitem{u0-fig}{5.5in} The function $j_{r0}({\bf u})$ for $Z=1$.  The
contours are equally spaced at intervals of $0.25$.

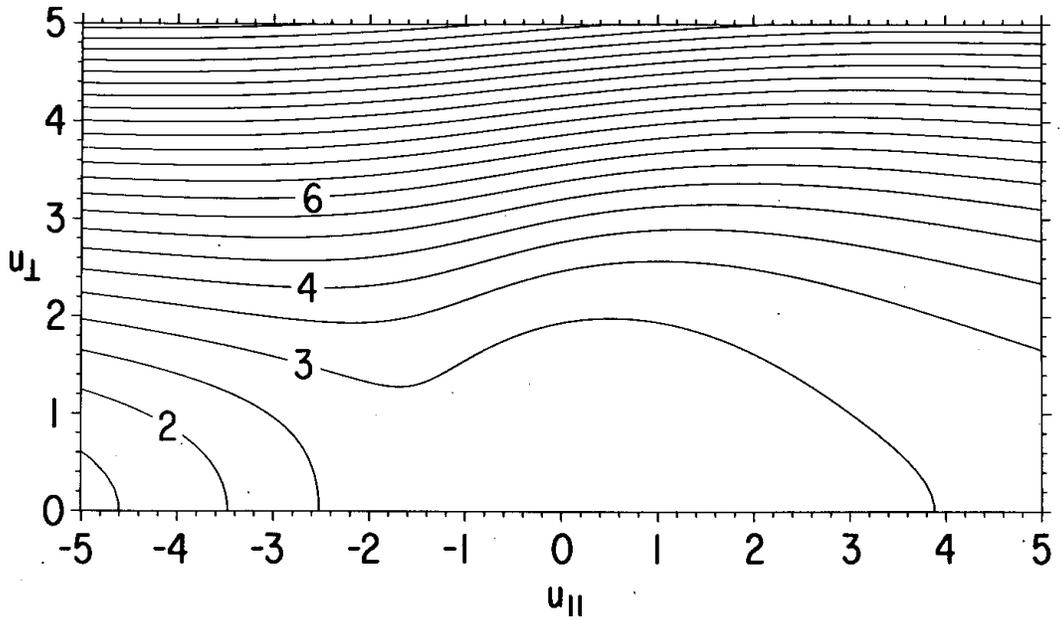
\figitem{E-fig}{5.5in} The perpendicular energy of the runaways ${\cal
E}_{\perp r}({\bf u})$ for $Z=1$.  The contours are equally spaced at
intervals of $0.5$.

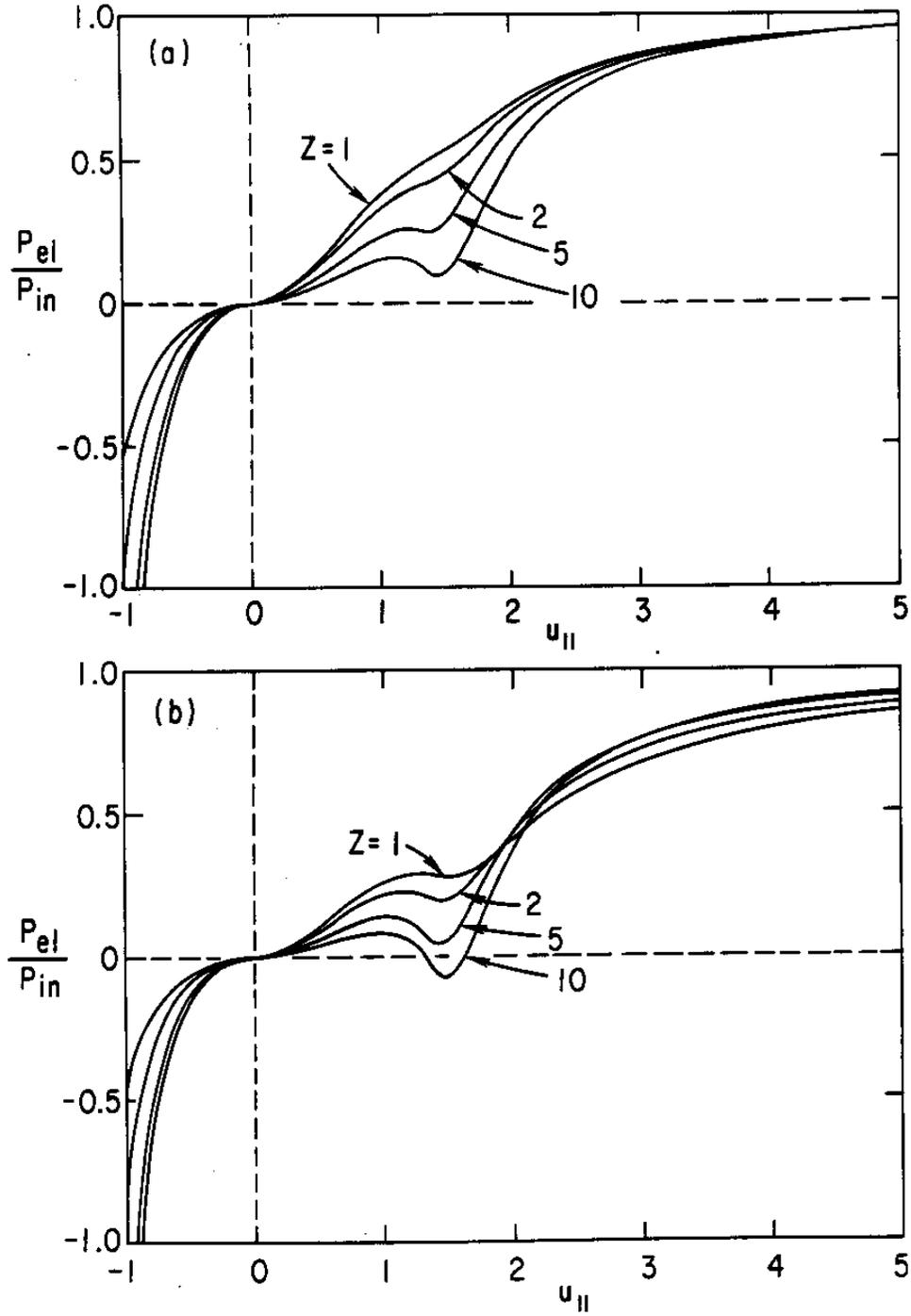
\figitem{W-figb}{5in} Efficiency for lower-hybrid current drive (a) and
for electron-cyclotron current drive (b) from Eqs.~(\ref{effa}).

\end{thefigures}
\end{document}